\documentclass[aps,prl,reprint,groupedaddress,showpacs,floatfix]{revtex4-1}

\usepackage[version=3]{mhchem} 
\usepackage{graphicx}
\usepackage{dcolumn}
\usepackage{bm}
\usepackage[mathlines]{lineno}
\usepackage{amsmath}
\usepackage{amssymb}
\usepackage{natbib}

\begin{document}

\title{Designing isoelectronic counterparts to layered group V semiconductors}

\author{Zhen Zhu}
\affiliation{Physics and Astronomy Department,
             Michigan State University,
             East Lansing, Michigan 48824, USA}

\author{Jie Guan}
\affiliation{Physics and Astronomy Department,
             Michigan State University,
             East Lansing, Michigan 48824, USA}

\author{Dan Liu}
\affiliation{Physics and Astronomy Department,
             Michigan State University,
             East Lansing, Michigan 48824, USA}

\author{David Tom\'{a}nek}
\email%
{tomanek@pa.msu.edu}%
\affiliation{Physics and Astronomy Department,
             Michigan State University,
             East Lansing, Michigan 48824, USA}
             



\begin{abstract}
In analogy to III-V compounds, which have significantly broadened
the scope of group IV semiconductors, we propose IV-VI compounds
as isoelectronic counterparts to layered group V semiconductors.
Using {\em ab initio} density functional theory, we study yet
unrealized structural phases of silicon mono-sulfide (SiS). We
find the black-phosphorus-like $\alpha$-SiS to be almost equally
stable as the blue-phosphorus-like $\beta$-SiS. Both $\alpha$-SiS
and $\beta$-SiS monolayers display a significant, indirect band
gap that depends sensitively on the in-layer strain. Unlike 2D
semiconductors of group V elements with the corresponding
nonplanar structure, different SiS allotropes show a strong
polarization either within or normal to the layers. We find that
SiS may form both lateral and vertical heterostructures with
phosphorene at a very small energy penalty, offering an
unprecedented tunability in structural and electronic properties
of SiS-P compounds.
\end{abstract}

\maketitle

\section{Introduction}

2D semiconductors of group V elements, including phosphorene and
arsenene, have been rapidly attracting interest due to their
significant fundamental band gap, large density of states near the
Fermi level, and high and anisotropic carrier
mobility\cite{{DT229},{DT230},{DT232},{Li2014}}. Combination of
these properties places these systems very favorably in the group
of contenders for 2D electronics applications beyond
graphene\cite{{PKimPRL07},{graphane}} and transition metal
dichalcogenides\cite{Kis2011}. Keeping in mind that the scope of
group IV semiconductors such as Si has been broadened
significantly by introducing isoelectronic III-V compounds, it is
intriguing to see, whether the same can be achieved in IV-VI
compounds that are isoelectronic to group V elemental
semiconductors. Even though this specific point of view has not
yet received attention, there has been interest in specific IV-VI
compounds, such as SnS, SnSe and GeTe, for thermoelectric and
photovoltaic
applications.\cite{{thermoelectric},{photovoltaic},{SnS-opto-expt}}
It appears likely that specific search for isoelectronic
counterparts of layered semiconductors such as phosphorene and
arsenene may guide us to yet unexplored 2D semiconducting IV-VI
compounds that are stable, flexible, and display a tunable band
gap.

\begin{figure*}[t]
\includegraphics[width=1.8\columnwidth]{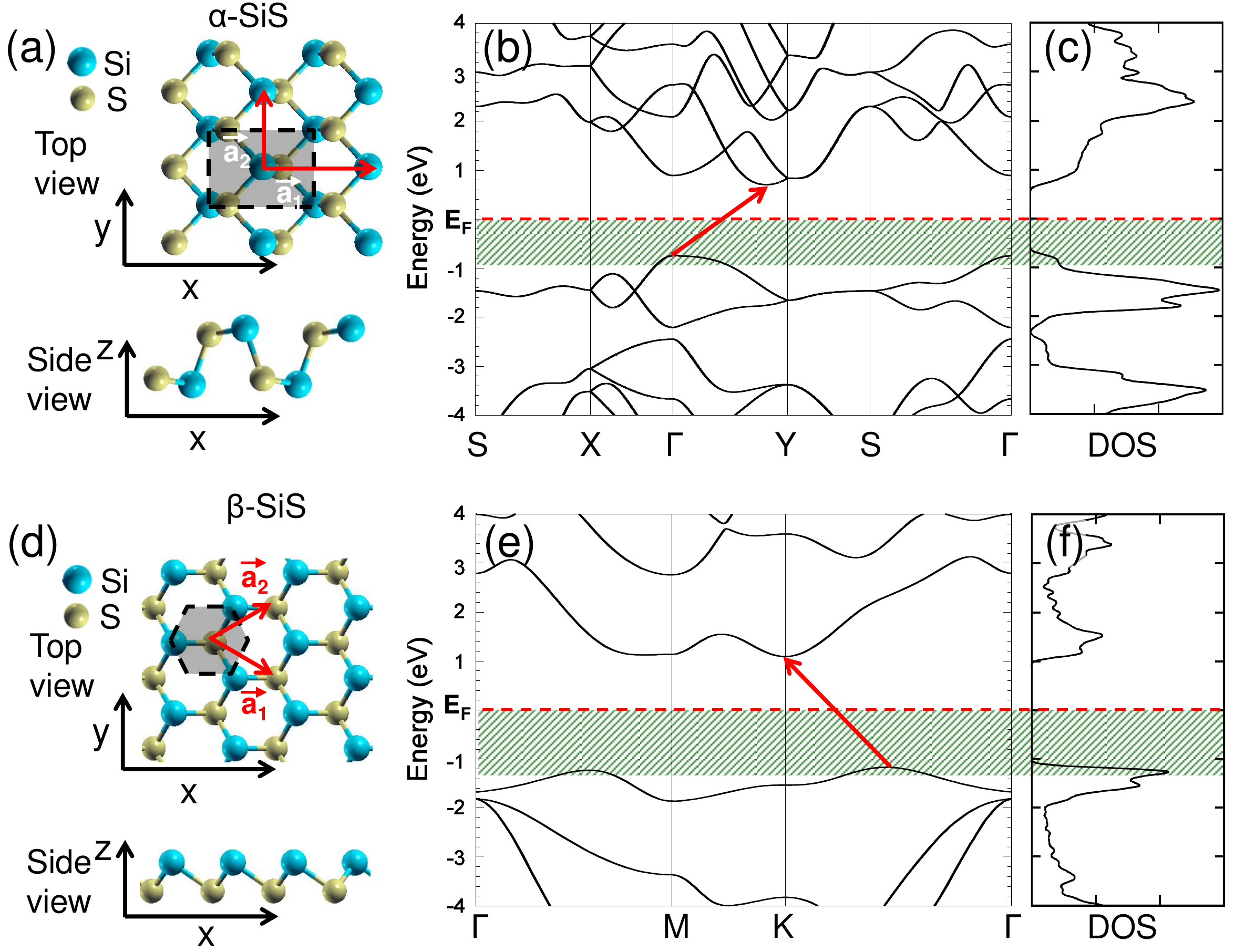}
\caption{(Color online) Atomic and electronic structure of [(a) to
(c)] $\alpha$-SiS and [(d) to (f)] $\beta$-SiS monolayers. [(a)
and (d)] Ball-and-stick models of the geometry, with S and Si
atoms distinguished by size and color and the Wigner-Seitz cell
indicated by the shaded region. [(b) and (e)] The electronic band
structure and [(c) and (f)] the electronic density of states (DOS)
of the systems. The energy range between $E_F$ and 0.2~eV below
the top of the valence band, indicated by the green shading, is
used to identify valence frontier states. \label{fig1}}
\end{figure*}


As a yet unexplored IV-VI compound, we study the layered structure
of silicon mono-sulfide (SiS). We use {\em ab initio} density
functional theory (DFT) to identify stable allotropes, determine
their equilibrium geometry and electronic structure. We have
identified two nearly equally stable allotropes, namely the
black-phosphorus-like $\alpha$-SiS and the blue-phosphorus-like
$\beta$-SiS, and show their structure in Fig.~\ref{fig1}(a) and
(d). Both $\alpha$-SiS and $\beta$-SiS monolayers display a
significant, indirect band gap that depends sensitively on the
in-layer strain. Unlike 2D semiconductors of group V elements with
the corresponding nonplanar structure, different SiS allotropes
show a strong polarization either within or normal to the layers.
We find that SiS may form both lateral and vertical
heterostructures with phosphorene at a very small energy penalty,
offering an unprecedented tunability in structural and electronic
properties of SiS-P compounds.

\section{Results and discussion}

\begin{figure}[t]
\includegraphics[width=1.0\columnwidth]{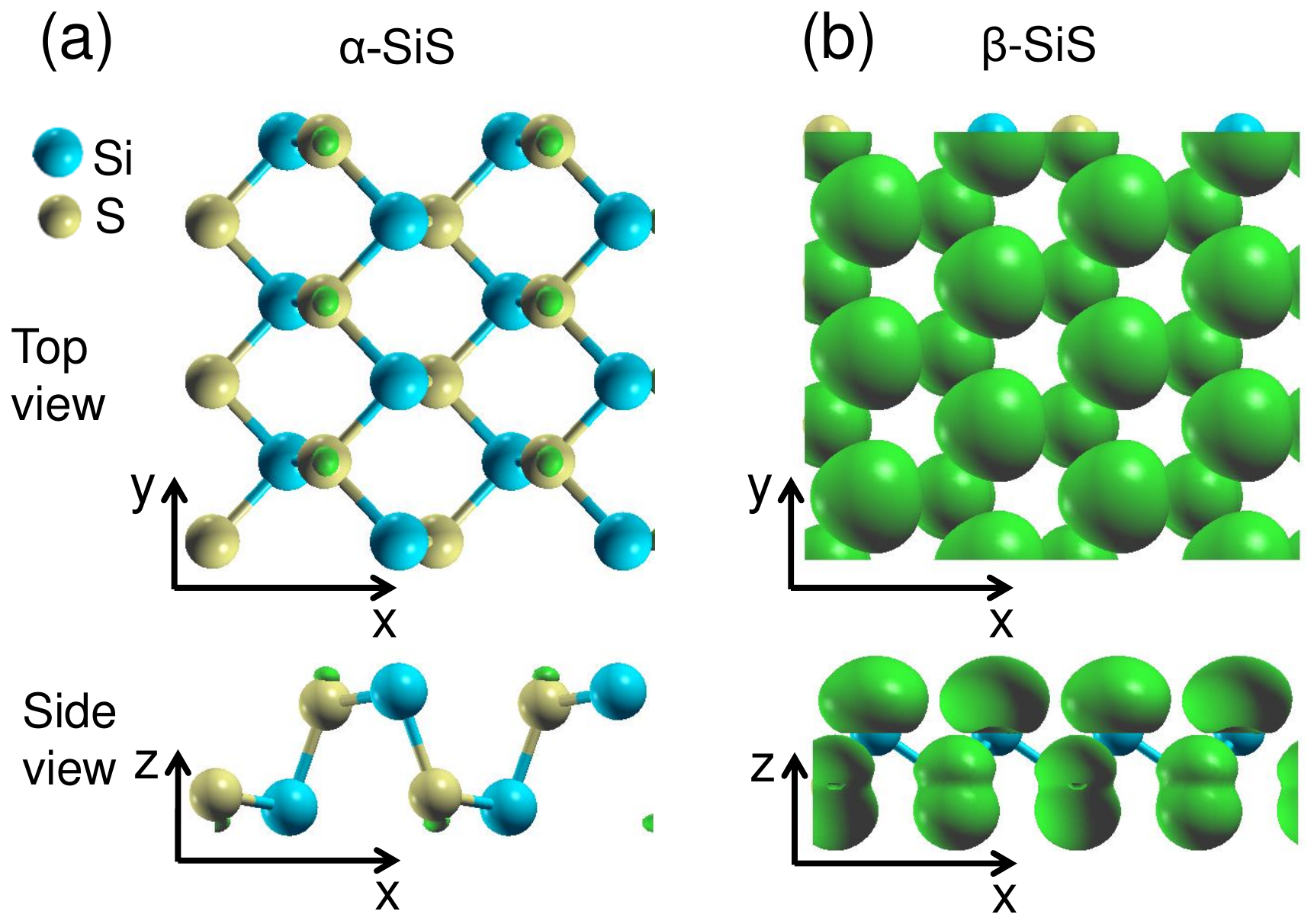}
\caption{(Color online) Electron density $\rho_{vb}$ associated
with states in the energy range between the Fermi level $E_F$ and
0.2~eV below the top of the valence band, shown by the shaded
region in Fig.~\ref{fig1}, in a monolayer of (a) $\alpha$-SiS and
(b) $\beta$-SiS. $\rho_{vb}=2.5{\times}10^{-3}$~e/{\AA}$^3$
contours are superposed with ball-and-stick models of related
structures. \label{fig2}}
\end{figure}


Since all atoms in $sp^3$ layered structures of group V elements
are threefold coordinated, the different allotropes can all be
topologically mapped onto the honeycomb lattice of graphene with 2
sites per unit cell.\cite{DT240} An easy way to generate IV-VI
compounds that are isoelectronic to group V monolayers is to
occupy one of these sites by a group IV and the other by a group
VI element. In this way, we have generated the orthorhombic
$\alpha$-SiS monolayer structure, shown in Fig.~\ref{fig1}(a),
from a monolayer of black phosphorene (or $\alpha$-P). The
hexagonal $\beta$-SiS monolayer, shown in Fig.~\ref{fig1}(d), has
been generated in the same way from the blue phosphorene (or
$\beta$-P) structure.

The monolayer structures have been optimized using DFT with the
Perdew-Burke-Ernzerhof (PBE)\cite{PBE} exchange-correlation
functional, as discussed in the Methods Section. We found that
presence of two elements with different local bonding preferences
increases the thickness of the SiS monolayers when compared to the
phosphorene counterparts. The 2D lattice of $\alpha$-SiS is
spanned by the orthogonal Bravais lattice vectors
$|\vec{a}_1|=4.76$~{\AA} and $|\vec{a}_2|=3.40$~{\AA}, which are
about 3\% longer than the lattice vectors of black phosphorene.
The 2D hexagonal lattice of $\beta$-SiS is spanned by two Bravais
lattice vectors
$a=|\vec{a}_1|=|\vec{a}_2|=3.35$~{\AA}, which are ${\lesssim}1$\%
longer than those of blue phosphorene. Unlike in the counterpart
structures of the phosphorene monolayer, we find $\beta$-SiS to be
more stable by about 12~meV/atom than $\alpha$-SiS. We checked the
phonon spectrum of free-standing $\alpha$-SiS and $\beta$-SiS
monolayers and found no soft modes that would cause a spontaneous
collapse of the structure.

An apparently different layered SiS structure had been
synthesized\cite{sis-exp-GS} by exposing the layered CaSi$_2$
structure to S$_2$Cl$_2$. It appears that the S$_2$Cl$_2$ reagent
reacted with the Ca atoms intercalated between silicene layers in
CaSi$_2$ by forming CaCl$_2$ and saturating the valencies of the
silicene layers by $-$S$-$S bridges. Even though no structural
information has been provided in that study, we identified a
likely candidate structure and found it to be less stable than the
postulated SiS allotropes, on which we focus next.

Since the electronegativity of S is higher than that of Si, we
expect an electron transfer from Si to S atoms. Based on a
Mulliken population analysis, we estimate a net transfer of 0.3
electrons in $\alpha$-SiS and 0.2 electrons in $\beta$-SiS from
silicon to sulfur atoms. This charge redistribution, combined with
the nonplanarity of the structure, causes a net polarization. We
find an in-plane polarization for $\alpha$-SiS and and an
out-of-plane polarization for $\beta$-SiS.


Whereas DFT generally provides an accurate description of the
total charge density and equilibrium geometry, interpretation of
Kohn-Sham energy eigenvalues as quasiparticle energies is more
problematic. Still, we present our DFT-PBE results for the
electronic structure of $\alpha$-SiS and $\beta$-SiS monolayers in
Fig.~\ref{fig1}. Even though the fundamental band gaps are
typically underestimated in this approach, the prediction that
both systems are indirect-gap semiconductors is likely correct.
Our calculated band structure and the corresponding density of
states for $\alpha$-SiS, presented in Fig.~\ref{fig1}(b,c),
suggest that the fundamental band gap value $E_g=1.44$~eV should
be significantly larger than in the isoelectronic black
phosphorene counterpart. The band structure near the top of the
valence band shows a significant anisotropy when comparing the
$\Gamma-X$ and $\Gamma-Y$ directions. In analogy to phosphorene,
$\alpha$-SiS should exhibit a higher hole mobility along the
$x$-direction than along the $y$-direction.

The DFT-based fundamental band gap $E_g=2.26$~eV in the monolayer
of $\beta$-SiS is even larger. The band structure in the symmetric
honeycomb lattice of $\beta$-SiS is rather isotropic, as seen in
Fig.~\ref{fig1}(e). The top of the valence band is very flat,
resulting in a heavy hole mass and a large density of states (DOS)
in that region, as seen in Fig.~\ref{fig1}(f).

\begin{figure*}[t]
\includegraphics[width=1.6\columnwidth]{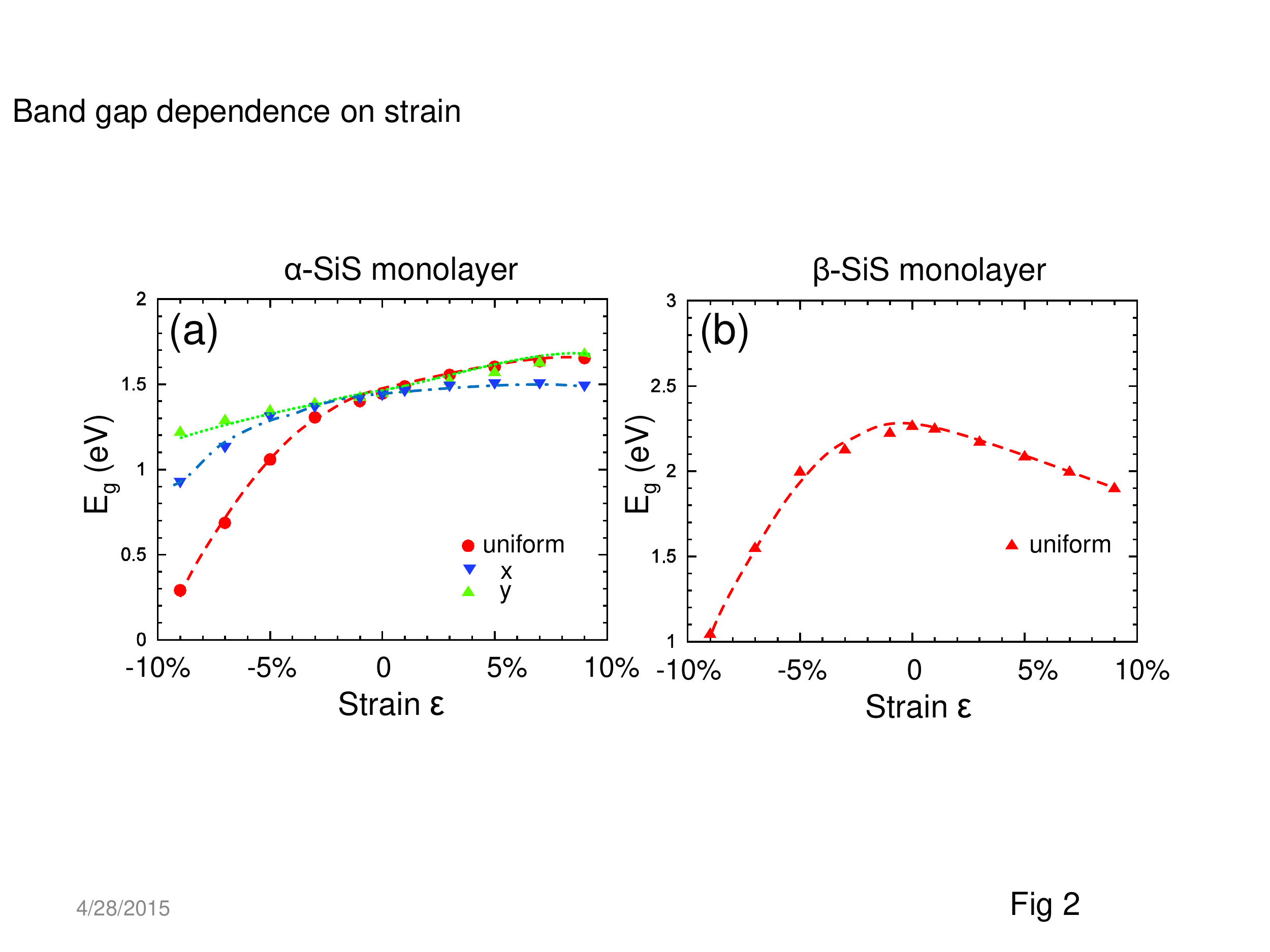}
\caption{(Color online) Electronic band gap of (a) $\alpha$-SiS
and (b) $\beta$-SiS monolayers as a function of the in-layer
strain. Besides the uniform strain, we also show results for
uniaxial strain along the $x$ and $y$ direction in the anisotropic
structure of $\alpha$-SiS in (a). The lines are guides to the eye.
\label{fig3}}
\end{figure*}

The character of frontier states is of interest not only for a
microscopic understanding of the conduction channels, but is also
crucial for the design of optimum contacts.\cite{DT243} Whereas
DFT-based band gaps are typically underestimated as mentioned
above, the electronic structure of the valence and the conduction
band region in DFT is believed to closely correspond to
experimental results. In Fig.~\ref{fig2}, we show the charge
density associated with frontier states near the top of the
valence band. These states, which correspond to the energy range
highlighted by the green shading in the band structure of
$\alpha$-SiS in Fig.~\ref{fig1}(b) and that of $\beta$-SiS in
Fig.~\ref{fig1}(e), cover the energy range between the mid-gap and
0.2~eV below the top of the valence band. The valence frontier
states of $\alpha$-SiS in Fig.~\ref{fig2}(a) and $\beta$-SiS in
Fig.~\ref{fig2}(b) are similar in spite of the notable charge
density differences caused by the larger DOS of $\beta$-SiS in
this energy range. The difference between S and Si atoms is also
reflected in the character of the valence frontier states at these
sites. Whereas $p_z$ states contribute most at sulfur sites,
silicon sites contribute a mixture of $s$ and $p_z$ states. These
frontier states differ from those of phosphorene, which are
related to lone pair electron states.

Similar to phosphorene, the fundamental band gap value of SiS also
depends sensitively on the in-layer strain, as seen in
Fig.~\ref{fig3}. Due to their nonplanarity, accordion-like
in-layer stretching or compression of SiS structures may be
achieved at little energy cost as shown in the Supporting
Information. The energy cost is particularly low for a deformation
along the soft $x$-direction, requiring ${\lesssim}20$~meV/atom to
induce a ${\pm}9$\% in-layer strain. We believe that in view of
the softness of the structure, similar strain values may be
achieved during epitaxial growth on particular incommensurate
substrates. We also note that tensile strain values such as these
have been achieved experimentally in suspended graphene membranes
that are much more resilient to
stretching due to their planar geometry and stronger bonds%
\cite{{AFMHone},{AFMMcEuen},{strainbending}}. Consequently, we
believe that strain engineering is a viable way to effectively
tune the fundamental band gap in these systems.

Our results for $\alpha$-SiS in Fig.~\ref{fig3}(a) indicate that
the band gap decreases when the structure is compressed and
increases slightly when it is stretched. The largest change in the
band gap, namely its reduction to 0.3~eV, may be achieved during a
$9$\% compression. As seen in Fig.~\ref{fig3}(b), we expect the
fundamental band gap of $\beta$-SiS to be reduced during both
stretching and compression. Within the ${\pm}9$\% range, we find
that the band gap may be tuned in the range from
${\approx}1.0-2.3$~eV. This high degree of band gap tunability in
SiS appears particularly attractive for potential applications in
flexible electronics.

\begin{figure*}[t]
\includegraphics[width=1.8\columnwidth]{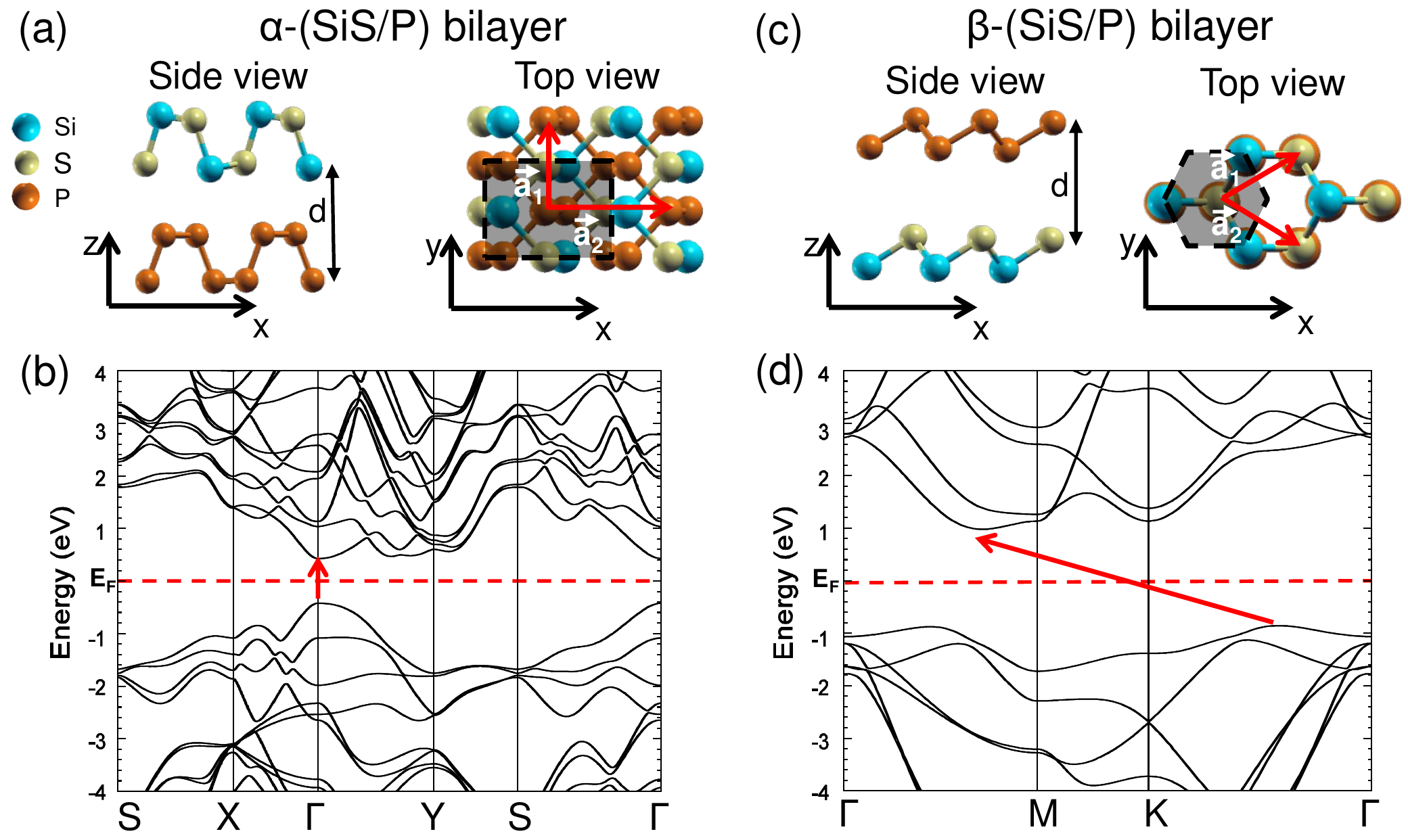}
\caption{(Color online) Optimum geometry and electronic band
structure of [(a) and (b)] an $\alpha$-(SiS/P) bilayer and [(c)
and (d)] a $\beta$-(SiS/P) bilayer. The optimum stacking of the
SiS and the phosphorene monolayers in the $\alpha$-(SiS/P) bilayer
in (a) is AB and that in the $\beta$-(SiS/P) bilayer is AA.
\label{fig4}}
\end{figure*}

\begin{figure*}[t]
\includegraphics[width=1.8\columnwidth]{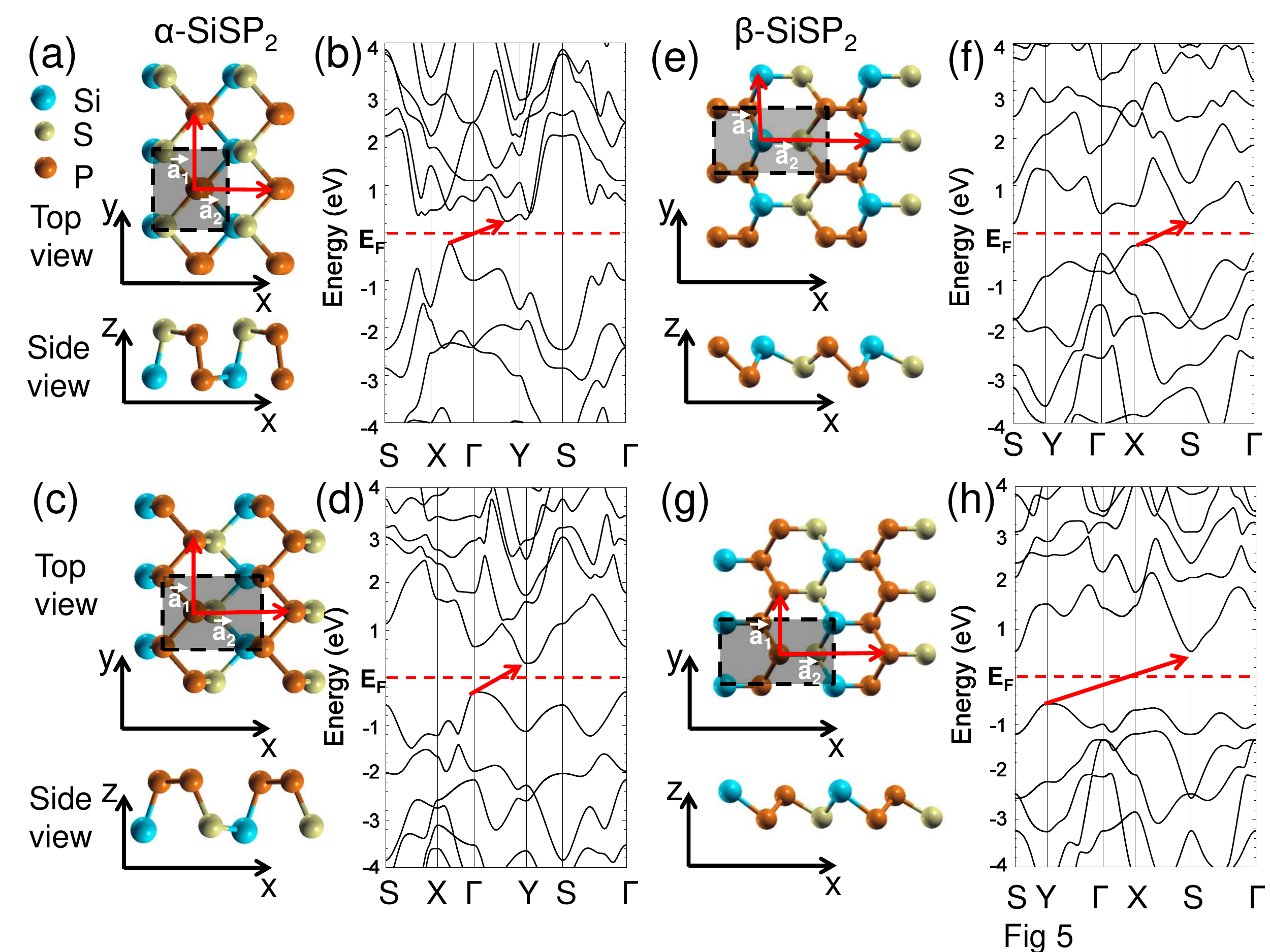}
\caption{(Color online) Optimum geometry and electronic band
structure of [(a) to (d)] two different in-layer heterostructures
of $\alpha$-SiSP$_2$ and [(e) to (h)] two different in-layer
heterostructures of $\beta$-SiSP$_2$. The heterostructures differ
in the arrangement of P atoms. \label{fig5}}
\end{figure*}

Since the geometry and lattice constants of SiS and phosphorene
are very similar, it is likely that the two could interface
naturally in lateral and vertical heterostructures, thus further
advancing the tunability of their electronic properties. In
Fig~\ref{fig4}, we present electronic structure results for
bilayers consisting of an SiS and a phosphorene monolayer in both
$\alpha$- and $\beta$-phases as the simplest examples of vertical
heterostructures.
We have optimized the bilayer structures assuming
commensurability, i.e. setting the primitive unit cells of each
monolayer to be the same. The optimum geometry of the
$\alpha$-(SiS/P) bilayer is shown in Fig~\ref{fig4}(a) and that of
the $\beta$-(SiS/P) bilayer in Fig~\ref{fig4}(b).

We find the inter-layer interaction in the two bilayer systems to
be rather weak, amounting to ${\lesssim}20$~meV/atom based on our
DFT-PBE calculations. Taking better account of van der Waals
interactions would likely increase the calculated interaction
energy and reduce the large interlayer separation of
$d{\lesssim}5.4$~{\AA}. Whereas the precise interlayer interaction
and separation are not of primary concern here, our most important
finding is that the weak interaction is not purely dispersive in
nature, since we find a substantial rehybridization of states
between the adjacent SiS and phosphorene layers. Consequently, as
we show more explicitly in the Supporting Information, the bilayer
band structure is not a mere superposition of the two monolayer
band structures in the same assumed geometry.

As shown in Fig.~\ref{fig4}(b), the $\alpha$-(SiS/P) bilayer is a
direct-gap semiconductor with $E_g{\approx}0.8$~eV, smaller than
the band gap in either isolated monolayer. Since the gap is
indirect in the $\alpha$-SiS monolayer, whereas it is direct in
$\alpha$-P, the cause for the direct band gap in $\alpha$-SiS/P
may be the dominance of P states near the gap. We find this
conjecture confirmed by examining the frontier orbitals of the
bilayer. As we show in the Supporting Information, the frontier
states in the conduction band region of the bilayer are
essentially purely P-based, whereas those in the valence band
region have only a small contribution from the SiS layer.

The electronic band structure of the AA-stacked $\beta$-(SiS/P)
bilayer is presented in Fig.~\ref{fig4}(d). The fundamental band
gap is indirect in this system and its value $E_g{\approx}1.7$~eV
is again smaller than in isolated blue phosphorene and $\beta$-SiS
monolayers. As seen in the Supporting Information, also the
frontier states of $\beta$-(SiS/P) are dominated by P orbitals
both in the valence and the conduction band region.

Since both SiS and phosphorene are rather flexible, they may
adjust to each other and form also in-layer heterostructures at
little or no energy penalty. We constructed two types of SiSP$_2$
heterostructures for both $\alpha$ and $\beta$ allotropes and show
their geometry and electronic structure in Fig.~\ref{fig5}. One
type of heterostructures, shown in Fig.~\ref{fig5}(a) and
\ref{fig5}(e), contains P-P and Si-S atom pairs completely
separated from like atom pairs. The other type of
heterostructures, presented in Fig.~\ref{fig5}(c) and
\ref{fig5}(g), contains alternating, contiguous SiS and phosphorus
chains. These structures maintain a rectangular lattice with four
atoms per unit cell. Generally, we found the in-layer
heterostructures to be less stable than pure phosphorene and SiS
monolayers. The least stable heterostructures among these are
those with isolated P-P or Si-S atom pairs, shown in
Fig.~\ref{fig5}(a) and \ref{fig5}(e), which are
${\approx}0.2$~eV/atom less stable than SiS and phosphorene
monolayers due to their highly frustrated geometries. The
heterostructures with contiguous phosphorus and SiS chains,
presented in Fig.~\ref{fig5}(c) and \ref{fig5}(g), may better
optimize the nearest-neighbor environment. This causes less
frustration, making these systems only ${\approx}0.1$~eV/atom less
stable than isolated SiS and phosphorene monolayers.

We found all four in-layer heterostructures to be indirect-gap
semiconductors. As in the vertical heterostructures, we found the
fundamental band gaps to be substantially smaller than in isolated
SiS and phosphorene monolayers. As seen in Fig.~\ref{fig5}(a,b)
and \ref{fig5}(e,f), the fundamental band gap $E_g$ is close to
0.5~eV in the less stable heterostructures with isolated Si-S and
P-P pairs. We found larger band gap values in the more stable
heterostructures with contiguous SiS and P chains, namely
$E_g{\approx}0.6$~eV in $\alpha$-SiSP$_2$ shown in
Fig.~\ref{fig5}(c,d) and $E_g{\approx}1.1$~eV in $\beta$-SiSP$_2$
shown in Fig.~\ref{fig5}(g,h), both nearly half the value of the
smaller band gap in either SiS or phosphorene monolayers. These
findings indicate an intriguing possibility of isoelectronic
doping as an effective way to tune the electronic properties of
SiSP$_n$ systems.

In view of the different successful synthesis approaches used to
form layered IV-VI compounds such as SiS\cite{sis-exp-GS} or
SnS\cite{sns-growth} with a structure similar to $\alpha$-SiS, we
believe that a suitable synthesis path will be found to form also
the allotropes introduced in this study.
The weak inter-layer interaction in layered pure SiS compounds or
SiS/P heterostructures should allow for a mechanical exfoliation
yielding monolayers and bilayers, same as in graphene and
phosphorene.
Chemical Vapor Deposition (CVD), which had been used successfully
in the past to grow graphene\cite{{KimNat2009},{ReinaNL2009}} and
silicene\cite{VogtPRL2012}, may become ultimately the most common
approach to grow few-layer SiS on specific substrates.

As suggested above, compounds that are isoelectronic to group V
layered semiconductors are not limited to pure IV-VI systems, but
may contain other group V elements in the same layer or in a
vertical few-layer heterostructure. The initial interest in black
phosphorene could thus be significantly expanded by considering
the whole range of group IV elements including Si, Ge, Sn, and Pb,
group VI elements such as S, Se, and Te, and group V elements such
as P, As, Sb and Bi, leading to a virtually limitless number of
compounds and structural phases. The large family of IV-VI systems
that are isoelectronic to group V elemental semiconductors will
include 2D semiconductors with a sizeable fundamental band gap, a
high carrier mobility and chemical stability. Expanding this
family to heterostructures containing also group V elements should
provide a way to effectively tune the electronic properties of the
pristine structures, which will likely bring unlimited richness to
the field of 2D semiconductors.

In conclusion, we have proposed IV-VI compounds as isoelectronic
counterparts to layered group V semiconductors in analogy to III-V
compounds, which have significantly broadened the scope of group
IV semiconductors. Using {\em ab initio} density functional
theory, we have identified yet unrealized structural phases of
silicon mono-sulfide (SiS) including the black-phosphorus-like
$\alpha$-SiS and the almost equally stable, blue-phosphorus-like
$\beta$-SiS. We found that both $\alpha$-SiS and $\beta$-SiS
monolayers display a significant, indirect band gap that depends
sensitively on the in-layer strain. Unlike 2D semiconductors of
group V elements with the corresponding nonplanar structure,
different SiS allotropes are polar. We found that SiS may form
both lateral and vertical heterostructures with phosphorene at a
very small energy penalty, offering an unprecedented tunability in
structural and electronic properties of SiS-P compounds. Combining
other group IV and group VI elements with group V elements is
expected to lead to a large family of layered semiconductors with
an unprecedented richness in structural and electronic properties.

\section{Methods}

Our computational approach to study the equilibrium structure,
stability and electronic properties of SiS structures is based on
{\em ab initio} density functional theory with the
Perdew-Burke-Ernzerhof (PBE)\cite{PBE} exchange-correlation
functional as implemented in the \textsc{SIESTA}\cite{SIESTA} and
VASP\cite{VASP,VASP2,VASP3,VASPPAW} codes. We use periodic
boundary conditions throughout the study, with few-layer
structures represented by a periodic array of slabs separated by a
vacuum region of ${\geq}15$~{\AA}.
In our \textsc{SIESTA} calculations we used norm-conserving
Troullier-Martins pseudopotentials\cite{Troullier91} and a
double-$\zeta$ basis including polarization orbitals. The
reciprocal space has been sampled by a fine
grid\cite{Monkhorst-Pack76} of $8{\times}8{\times}1$~$k$-points in
the Brillouin zone of the primitive unit cell of the 2D
structures.
We used a mesh cutoff energy of $180$~Ry to determine the
self-consistent charge density, which provided us with a precision
in total energy of ${\leq}2$~meV/atom. In \textsc{VASP}
calculations, we used an energy cut-off of 500~eV and the same
k-point sampling as mentioned above. All geometries have been
optimized using the conjugate gradient method\cite{CGmethod},
until none of the residual Hellmann-Feynman forces exceeded
$10^{-2}$~eV/{\AA}. Equilibrium structures and energies based on
\textsc{SIESTA} have been checked against values based on the
{\textsc VASP} code.

\section{Acknoledgement}
This study has been supported by the National Science Foundation
Cooperative Agreement \#EEC-0832785, titled ``NSEC: Center for
High-rate Nanomanufacturing''. Computational resources have been
provided by the Michigan State University High Performance
Computing Center.


\providecommand{\latin}[1]{#1}
\providecommand*\mcitethebibliography{\thebibliography} \csname
@ifundefined\endcsname{endmcitethebibliography}
  {\let\endmcitethebibliography\endthebibliography}{}

\end{document}